\documentclass[twocolumn,reprint,
superscriptaddress,
  aps,
  pre,
  amsmath,amssymb,
  reprint, longbibliography
]{revtex4-1}
\usepackage{graphicx,color}
\usepackage{amsmath}
\usepackage{natbib}
\usepackage{epsfig}
\usepackage{epstopdf}
\usepackage{hyperref}
\begin{document}

\title{When do soft spheres become hard spheres?}

\author{Sergey Khrapak}
\email{Sergey.Khrapak@gmx.de}
\affiliation{Joint Institute for High Temperatures, Russian Academy of Sciences, 125412 Moscow, Russia}
\affiliation{Bauman Moscow State Technical University, 105005 Moscow, Russia}
\affiliation{Institut f\"ur Materialphysik im Weltraum, Deutsches Zentrum f\"ur Luft- und Raumfahrt (DLR), 82234 We{\ss}ling, Germany}
\author{Nikita P. Kryuchkov}
\affiliation{Bauman Moscow State Technical University, 105005 Moscow, Russia}

\author{Lukiya A. Mistryukova}
\affiliation{Bauman Moscow State Technical University, 105005 Moscow, Russia}

\author{Stanislav O. Yurchenko}
\email{st.yurchenko@mail.ru}
\affiliation {Bauman Moscow State Technical University, 105005 Moscow, Russia}

\begin{abstract}
The conventional (Zwanzig-Mountain) expressions for instantaneous elastic moduli of simple fluids predict their divergence as the limit of hard sphere (HS) interaction is approached.
However, elastic moduli of a true HS fluid are finite. Here we demonstrate that this paradox reveals the soft- to hard-sphere crossover in fluid excitations and thermodynamics. With extensive \emph{in-silico} study of fluids with repulsive power-law interactions ($\propto r^{-n}$), we locate the crossover at $n\simeq 10-20$ and develop a simple and accurate model for the HS regime. The results open novel prospects to deal with the elasticity and related phenomena in various systems, from simple fluids to melts and glasses.

\end{abstract}

\date{\today}

\maketitle

\section{Introduction}

Understanding the mechanisms governing elastic moduli of substances is an important  problem~\cite{MasonPRL1995,KokuboBioMat2003,MasonRA2000,DyreRMP2006,NemilovJNCS2006}:
Elastic moduli are directly related to long-wavelength (sound) excitations -- phonons, which play a crucial role in condensed matter, materials science, and soft matter.

For example, the celebrated Lindemann melting criterion states~\cite{Lindemann} that a three-dimensional (3D) solid melts when the vibrational amplitude of atoms around their equilibrium positions reaches about $\sim 0.1$ of the interatomic distance. Since the vibrational amplitude is dominated by long-wavelength excitations, the melting temperature can be expressed in terms of the shear and bulk moduli~\cite{BuchenauPRE2014, KhrapakPRR2020}. Another example is the Berezinskii-Kosterlitz-Thouless-Halperin-Nelson-Young (BKTHNY) theory of two-dimensional (2D) melting~\cite{KosterlitzJPC1973, NelsonPRB1979, YoungPRB1979, BerezinskiiJETP1971, KosterlitzRMP2017, Ryzhov2017}:
The condition for dislocation unbinding, responsible for crystal melting, can be expressed in terms of the (2D) shear and bulk moduli. Additionally, there is a possibility to formulate a 2D  Lindemann-like criterion and relate it to the BKTHNY mechanism~\cite{KhrapakJCP2018, KhrapakPRR2020}.
The instantaneous bulk and shear moduli are related to the alpha-relaxation time in the framework of the shoving model, thus, playing an important role in the physics of glass-forming liquids~\cite{DyrePRE2004, DyreJCP2012}. For instance, the temperature dependence of the shear viscosity can be expressed via the instantaneous shear modulus~\cite{ChevallardPRR2020}. However, the effect of interaction softness on elastic moduli and collective excitations of fluids is still poorly understood.

The behaviour of elastic moduli in systems with steeply repulsive interactions has been remaining a rather controversial issue for the last 50 years.
The conventional (Zwanzig-Mountain) expressions for the high-frequency (instantaneous) bulk and shear moduli~\cite{ZwanzigJCP1965, Schofield1966} predict their \emph{divergence} as the hard-sphere (HS) limit is approached from the side of soft interactions~\cite{Frisch1966, HeyesJCP1994}.
However, this divergence is inconsistent with other observations.
The shear and bulk moduli of a true \emph{HS fluid} are non-singular and well defined~\cite{MillerJCP1969}, as well as elastic moduli of \emph{HS solids} ~\cite{FrenkelPRL1987,RungePRA1987,LairdJCP1992} and \emph{HS glass}~\cite{LowenJPCM1990}.
Finite values of the bulk modulus follow from finite isothermal and adiabatic sound velocities (evaluated from an appropriate equation of state)~\cite{RosenfeldJPCM1999}.
Finally, the finite shear modulus emerges from the analysis of transverse excitations in fluids~\cite{BrykJCP2017}.

The origin of this ``paradoxical'' situation has been attributed to the assumption of no structural relaxation upon density change~\cite{KhrapakSciRep2017,KhrapakPRE2019}.
The latter is well justified for soft interactions, but becomes unsuitable in the HS limit, because of an intrinsic length scale -- the HS diameter.
Hence, the divergence of the elastic moduli is artificial and the conventional expressions are just inappropriate in the HS limit. However, the question regarding where exactly the soft interaction stops to be soft enough to use the Zwanzig-Mountain expressions and how the HS limit is approached has remained obscure.

In this paper, we report on theoretical and extensive \emph{in silico} studies of crossover from the soft-sphere (SSp) to the HS limit.
Using molecular dynamic (MD) simulations, we consider a system of particles interacting via the inverse-power-law (IPL) pair repulsion, $\varphi(r)=\epsilon(\sigma/r)^n$, where $\epsilon$ and $\sigma$ are the energy and length scales, and $n$ is the IPL exponent.
With the methods developed in ~\cite{KryuchkovSciRep2019},
we analyse in detail excitation spectra in dense IPL fluids with $n=8,12,20,50,100,200$. We consistently compare the sound velocities obtained \emph{in silico} and theoretically using the SSp and HS models, and discover that the crossover from the SSp to the HS dynamics occurs in the range $n\simeq 10-20$. For larger $n$ a new simple theoretical approach is shown to provide good accuracy in estimating the elastic moduli.


\section{MD simulations}

\begin{figure}[!t]
\includegraphics[width=80mm]{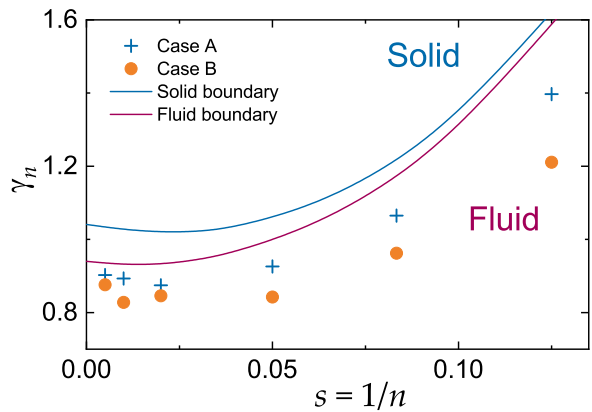}
\caption{Phase diagram of the IPL system: The fluid-solid coexistence region is bounded by the solid curves~\cite{AgrawalPRL1995,AgrawalMolPhys1995}; symbols denote the fluid state points studied in this work.}
\label{HSL-Fig1}
\end{figure}

The equilibrium state of the IPL system is determined by a single reduced parameter~\cite{DubinPRB1994}, $\gamma_n=\rho\sigma^3(\epsilon/T)^{3/n}$, where $\rho$ is the particle density and $T$ is the temperature (in energy units). The phase diagram of the IPL family is sketched in Fig.~\ref{HSL-Fig1}, using available data~\cite{AgrawalPRL1995,AgrawalMolPhys1995}.
In the HS limit, 
the phase state is determined by the packing fraction, $\eta=(\pi/6)\rho\sigma^3$, with the freezing (melting) point at $\eta_{f}\simeq 0.494$ ($\eta_m\simeq 0.545$)~\cite{BerthierRMP2011}.
The symbols in Fig.~\ref{HSL-Fig1} (cases A and B) denote the state points studied in this work.

We performed MD simulations of $N=10648$ particles in the canonical ($NVT$) ensemble with Nos{\'e}-Hoover thermostat and periodic boundary conditions in three dimensions.
We used dimensionless units of energy, length, and mass ($\epsilon=1$, $\sigma=1$, and $m=1$), a cutoff radius $r_c=7.5$, and a numerical time step $\Delta t=5\times 10^{-3}\sqrt{0.4 /T}$.
All simulations were performed for $2\times10^6$ time steps, using the LAMMPS package~\cite{Plimpton1995}.
The first $5\times10^5$ steps were used for equilibration and the following steps for the analysis.

The excitation spectra were obtained using the procedures described in Refs.~\cite{Yurchenko2018, KryuchkovSciRep2019, KryuchkovJPCL2019, YakovlevJPCL2020, KryuchkovPRL2020}.
First, we calculated the velocity current spectra~\cite{HansenBook, KryuchkovSciRep2019}:
\begin{equation}
\label{HSL-eq1}
 C_{L,T}(\mathbf{q},\omega) =\int dt\, e^{i\omega t} \mathrm{Re} \left<j_{L,T}(\mathbf{q},t)j_{L,T}(-\mathbf{q},0) \right>,
\end{equation}
where $\mathbf{q}$ and $\omega$ are the wave-vector and the frequency,
$\mathbf{j}_{L}=\mathbf{q}(\mathbf{q} \cdot \mathbf{j})/q^2$ and $\mathbf{j}_{T}=\mathbf{j} -\mathbf{j}_{L}$ are the longitudinal and transverse components of the current $\mathbf{j}(\mathbf{q},t)=N^{-1} \sum_s \mathbf{v}_s(t)\exp(i \mathbf{q} \mathbf{r}_s(t))$;
$\mathbf{v}_s(t)= \dot{\mathbf{r}}_s(t)$ is the velocity of $s$-th particle.
In isotropic fluids the directional dependence vanishes,  $C_{L,T}(\mathbf{q},\omega)\equiv C_{L,T}(q,\omega)$.
The total current spectra $C(q,\omega)=C_{L}(q,\omega)+2C_{T}(q,\omega)$ were fitted at each $q$-value with the two-oscillator model~\cite{KryuchkovSciRep2019}:
\begin{equation}
\label{HSL-eq2}
\begin{split}
 & C(q,\omega)\propto \frac{\Gamma_{\mathrm{hf}}}{(\omega-\omega_{\mathrm{hf}})^2+\Gamma_{\mathrm{hf}}^2}+
 \frac{\Gamma_{\mathrm{hf}}}{(\omega+\omega_{\mathrm{hf}})^2+\Gamma_{\mathrm{hf}}^2}+ \\ &
 \qquad\qquad +\frac{2\Gamma_{\mathrm{lf}}}{(\omega-\omega_{\mathrm{lf}})^2+\Gamma_{\mathrm{lf}}^2}+
 \frac{2\Gamma_{\mathrm{lf}}}{(\omega+\omega_{\mathrm{lf}})^2+\Gamma_{\mathrm{lf}}^2},
 \end{split}
\end{equation}
where $\omega_{\mathrm{hf}, \mathrm{lf}}$ and $\Gamma_{\mathrm{hf}, \mathrm{lf}}$ are the frequencies and damping rates of the high- and low-frequency branches. 

\begin{figure*}
\includegraphics[width=160mm]{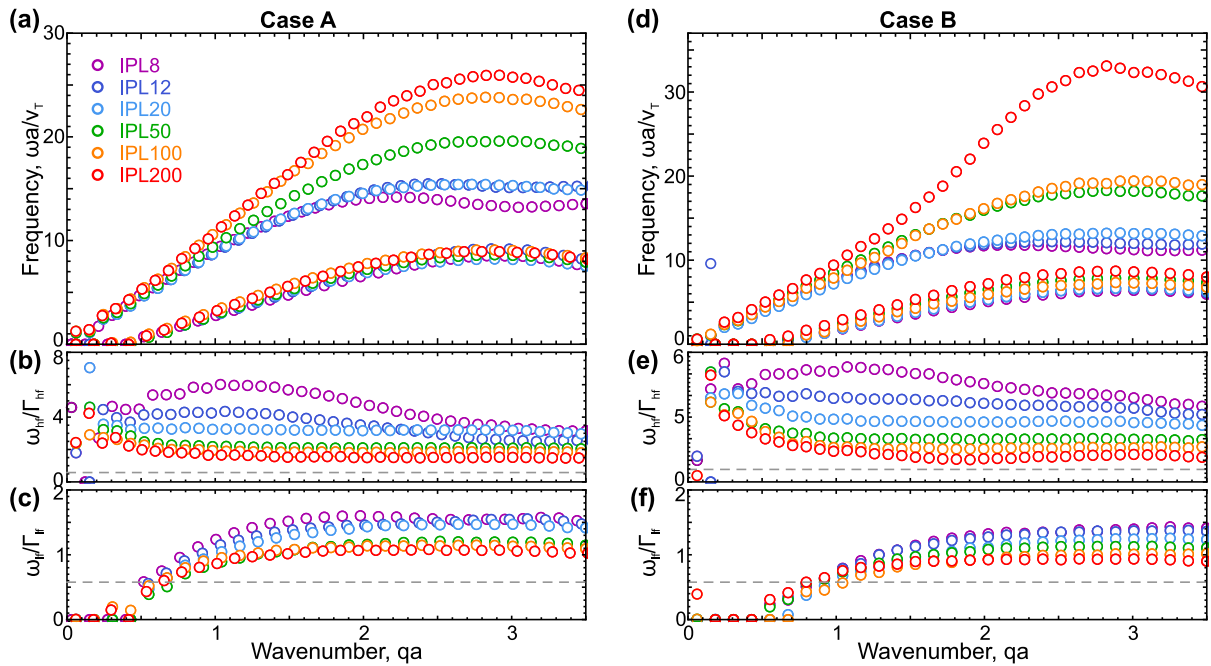}
\caption{
Collective excitation spectra in dense IPL fluids:
The results are shown for state points of Cases A (a)-(c) and B (d)-(f) in Fig.~\ref{HSL-Fig1}. The high- and low-frequency branches in (a) and (d) correspond to the longitudinal and transverse modes in the long-wavelength limit. Here $v_T=\sqrt{T/m}$ is the thermal velocity and $a=(4\pi\rho/3)^{-1/3}$ is the Wigner-Seitz radius.
The panels (b),(c) and (e),(f) show the ratios of the oscillation frequency to the damping rate for the high- and low-frequency branches. 
Below the grey dashed lines $\omega/\Gamma = 1/\sqrt{3}$, the branches become overdamped \cite{KryuchkovSciRep2019}.}
\label{HSL-Fig2}
\end{figure*}

\section{Results}

The excitation spectra in fluids are shown in Fig.~\ref{HSL-Fig2}~(a)-(c) and \ref{HSL-Fig2}~(d)-(f), for the state points of Cases A and B in Fig.~\ref{HSL-Fig1}. 
The low- and high-frequency dispersion curves shown in Figs.~\ref{HSL-Fig2}~(a) and \ref{HSL-Fig2}~(d)
behave similarly to those reported in Refs.~\cite{KryuchkovSciRep2019, KryuchkovJPCL2019, YakovlevJPCL2020, KryuchkovPRL2020}.
In the long-wavelength limit, as usual, these branches are attributed to the longitudinal and transverse collective modes \cite{KryuchkovSciRep2019, KryuchkovPRL2020}.
Remarkably, the branch $\omega_{\mathrm{lf}}$ behaves \emph{quasi-universally} (in reduced units), being almost independent of the IPL exponent, in contrast to $\omega_{\mathrm{hf}}$.
Here, the reduced maximum frequency increases with $n$, demonstrating clear positive sound dispersion for $n\gtrsim 20$.


In Figs.~\ref{HSL-Fig2} (b), (c), (e), and (f),
the ratio $\omega/\Gamma$ of the real frequency to the damping rate for the high- and low-frequency branches are shown to illustrate the quality factor of the oscillations.
The horizontal dashed lines correspond to $\omega/\Gamma = 1/\sqrt{3}$, below which even oscillating modes with $\omega \neq 0$ become overdamped \cite{KryuchkovSciRep2019, Yurchenko2018}; $\omega/\Gamma = 0$ corresponds to the gapped excitations of the low-frequency (transverse) branch.
The ratio $\omega/\Gamma$ grows monotonically for transverse excitations in the intermediate $q$-regime and is systematically larger for softer potentials (smaller $n$). On the contrary, the factor $\omega/\Gamma$ of the high-frequency branch has \emph{a maximum} in the vicinity of $qa\simeq 1.1$ in the SSp regime (the position of the maximum correlates with the transition from the hydrodynamic regime to the individual particles limit~\cite{KryuchkovSciRep2019}), but  drops \emph{monotonously} with $qa$ in the HS regime ($n \gtrsim 20$).
\emph{The qualitative changes} in the excitation spectra at $n\simeq 20$, highlighted in Fig.~\ref{HSL-Fig2}, point to a crossover from soft- to hard-sphere fluid collective dynamics, revealed here for the first time, to our knowledge.

To analyse this crossover in detail, consider the long-wavelength excitations in Fig.~\ref{HSL-Fig2}.
The longitudinal (bulk) and transverse (shear) sound velocities are evaluated from
\begin{equation}
\label{HSL-eq3}
c_s =\lim_{q\rightarrow 0}\frac{\partial \omega_{\mathrm{hf}}}{\partial q}, \quad\quad c_t =\lim_{q\rightarrow q_g+0}\frac{\partial \omega_{\mathrm{lf}}}{\partial q},
\end{equation}
where $c_t$ is calculated near the gap in reciprocal space (\emph{q-gap}), corresponding to the minimum wave number $q_g$, below which $\omega_{\mathrm{lf}}=0$, and where typically $\omega_{\mathrm{lf}} \simeq c_t(q-q_g)$~\cite{KryuchkovSciRep2019}.
With the linear fitting \eqref{HSL-eq3} of the MD excitation spectra (Fig.~\ref{HSL-Fig2}), we have obtained the sound velocities.


The sound velocities deduced from MD simulation contain information about the instantaneous bulk and shear moduli~\cite{KhrapakSciRep2017} via the relations $K_{\infty} =m\rho c_s^2$ and $G_{\infty} = m\rho c_t^2$, respectively. For repulsive interactions, including the HS limit, the instantaneous and adiabatic moduli are numerically close~\cite{KhrapakSciRep2017,KhrapakPRE2019,
KhrapakPoP2016_Relations}, and we do not distinguish between tham in the following. 
In the conventional SSp paradigm, the moduli are expressed via the pair potential $\varphi (r)$ and the RDF $g(r)$~\cite{ZwanzigJCP1965,Schofield1966}:
\begin{equation}
\label{HSL-eq4}
\begin{split}
& G_{\infty}=\rho T+\frac{2\pi \rho^2}{15}\int_0^{\infty}dr \; r^3 g(r)\left[r \varphi''(r)+4\varphi'(r)\right], \\
& K_{\infty}=\frac{5}{3}\rho T+\frac{2\pi \rho^2}{9}\int_0^{\infty}dr\;r^3g(r)\left[r \varphi''(r)-2\varphi'(r)\right].
\end{split}
\end{equation}
The first terms in expressions for $G_{\infty}$ and $K_{\infty}$ correspond to the kinetic (ideal gas) contribution, while the second ones are the configurational (excess) contribution, $G_{\mathrm{ex}}$ and $K_{\mathrm{ex}}$, which are dominant in dense fluids. In IPL fluids
$G_{\mathrm{ex}}$ and $K_{\mathrm{ex}}$ are directly related to the excess pressure $P_{\mathrm{ex}}$ as $G_{\mathrm{ex}}=(n-3)P_{\mathrm{ex}}/5$ and $K_{\mathrm{ex}}=(n+3)P_{\mathrm{ex}}/3$.

The paradoxical divergence of elastic moduli now becomes particularly clear:
As $n$ increases, $G_{\mathrm{ex}}$ and $K_{\mathrm{ex}}$ diverge as $\propto n$, because $P_{\mathrm{ex}}$ remains finite in the HS limit.
The spectra in Fig.~\ref{HSL-Fig2}, however, evidence that this is not the case and the actual elastic moduli ($\propto c_s^2$ and $c_t^2$) are finite.
This proves that the SSp expressions \eqref{HSL-eq4} become unsuitable at large $n$. But \emph{where exactly} do soft spheres become not so soft any more?

To answer the question, we have constructed a simple HS asymptotic model for elastic moduli at large $n$. We start with the Carnahan-Starling (CS) equation of state~\cite{CarnahanJCP1969} of the HS fluid. The pressure is written as
\begin{equation}
\label{HSL-eq5}
P(\rho,T)=\rho TZ(\eta), \qquad Z(\eta)=\frac{1+\eta+\eta^2-\eta^3}{(1-\eta)^3},
\end{equation}
where $Z(\eta)$ is the CS compressibility, and the effective HS packing fraction $\eta$ depends on the effective HS diameter $d_{\mathrm{eff}}$ specified below. The adiabatic bulk modulus $K_s$ follows straightforwardly from the factor $Z$~\cite{RosenfeldJPCM1999,KhrapakJCP2016}:
\begin{equation}
\label{HSL-eq6}
K_{s}=\rho T\left[Z(\eta)+\eta dZ(\eta)/d\eta+\tfrac{2}{3}Z^2(\eta)\right].
\end{equation}
For the shear modulus, we combine the expression for $G_{\infty}$ derived by Miller~\cite{MillerJCP1969,KhrapakPRE2019} with the approximation for $g'(1)$ reported in Ref.~\cite{TaoPRA1992}:
\begin{equation}
\label{HSL-eq7}
G_{\infty}= \rho T\left[1-\frac{8}{5}\eta g'(1)\right], \qquad g'(1)=-\frac{9\eta(1+\eta)}{2(1-\eta)^4},
\end{equation}
where $g'(1)$ denotes the reduced derivative at contact, $g'(1)=\lim_{\epsilon \rightarrow 0}\left[dg(x)/dx\right]_{x=1+\epsilon}$ with $x=r/d$.

The final step is to determine the effective HS diameter $d_{\rm eff}$. Several approximations have been proposed over the years, among the most familiar are those by Rowlinson; Barker and Henderson; and Stillinger~\cite{Rowlinson1964,Barker1967,Henderson2010,
Stillinger1976}. For the IPL potential they
result in the generic condition 
\begin{equation}
\label{HSL-eq8}
d_{\mathrm{eff}}={\mathcal A}\sigma\left(\epsilon/T\right)^{1/n},
\end{equation}
where ${\mathcal A}$ is a numerical factor, which tends to unity as $n\rightarrow \infty$, but appears  approximation-dependent at finite $n$. In the considered case the Rowlinson and Barker and Henderson (RBH) approximations coincide and we get $\mathcal A= \Gamma(1-1/n)$. Stillinger approximation yields $\mathcal A= (1/\ln2)^{1/n}$. Yet another approximation, which appears particularly suitable for the problem at hand, results from equating the potential interaction energy at $d_{\mathrm{eff}}$ to the average kinetic energy $3T/2$, resulting in $\mathcal A= (2/3)^{1/n}$. For a given effective HS diameter, the sound velocities are evaluated from $c_s = \sqrt{K_s/\rho m}$ and $c_t = \sqrt{G_{\infty}/\rho m}$ using Eqs.~(\ref{HSL-eq6}) and (\ref{HSL-eq7}).

\begin{figure}[!t]
\includegraphics[width=80mm]{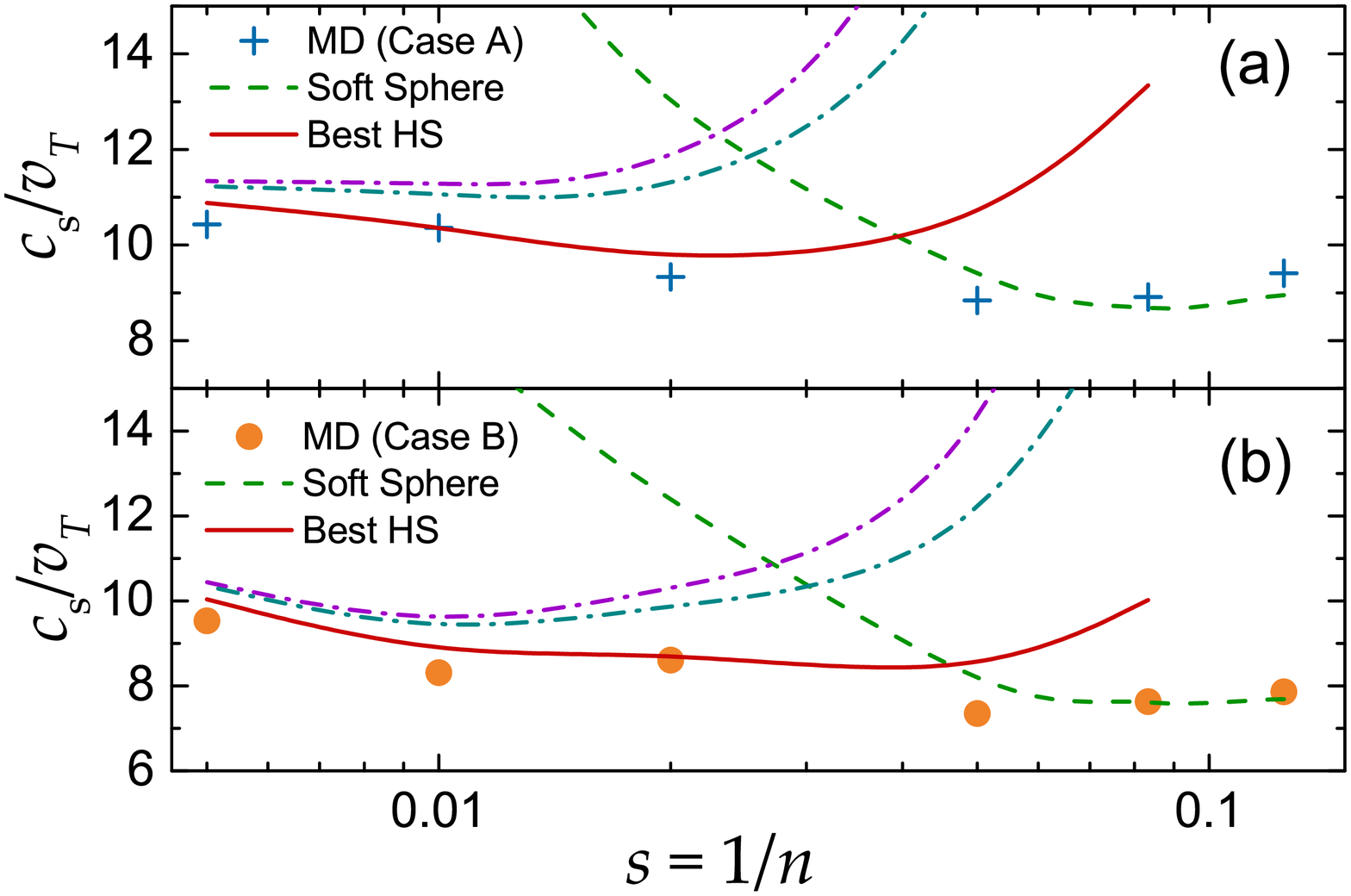}
\caption{Reduced longitudinal sound velocity $c_s/v_{\mathrm{T}}$ versus the softness parameter $s=1/n$ of the IPL fluids for Case A (a) and Case B (b). Symbols correspond to results from MD simulation. The dashed and solid curves correspond to the SSp and the best HS asymptote, \eqref{HSL-eq4} and \eqref{HSL-eq6}, respectively. Two additional dash-dotted curves in the HS regime correspond to the effective HS diameters estimated using RBH (upper) and Stillinger (lower) approaches.}
\label{HSL-Fig3}
\end{figure}

\begin{figure}[!t]
\includegraphics[width=80mm]{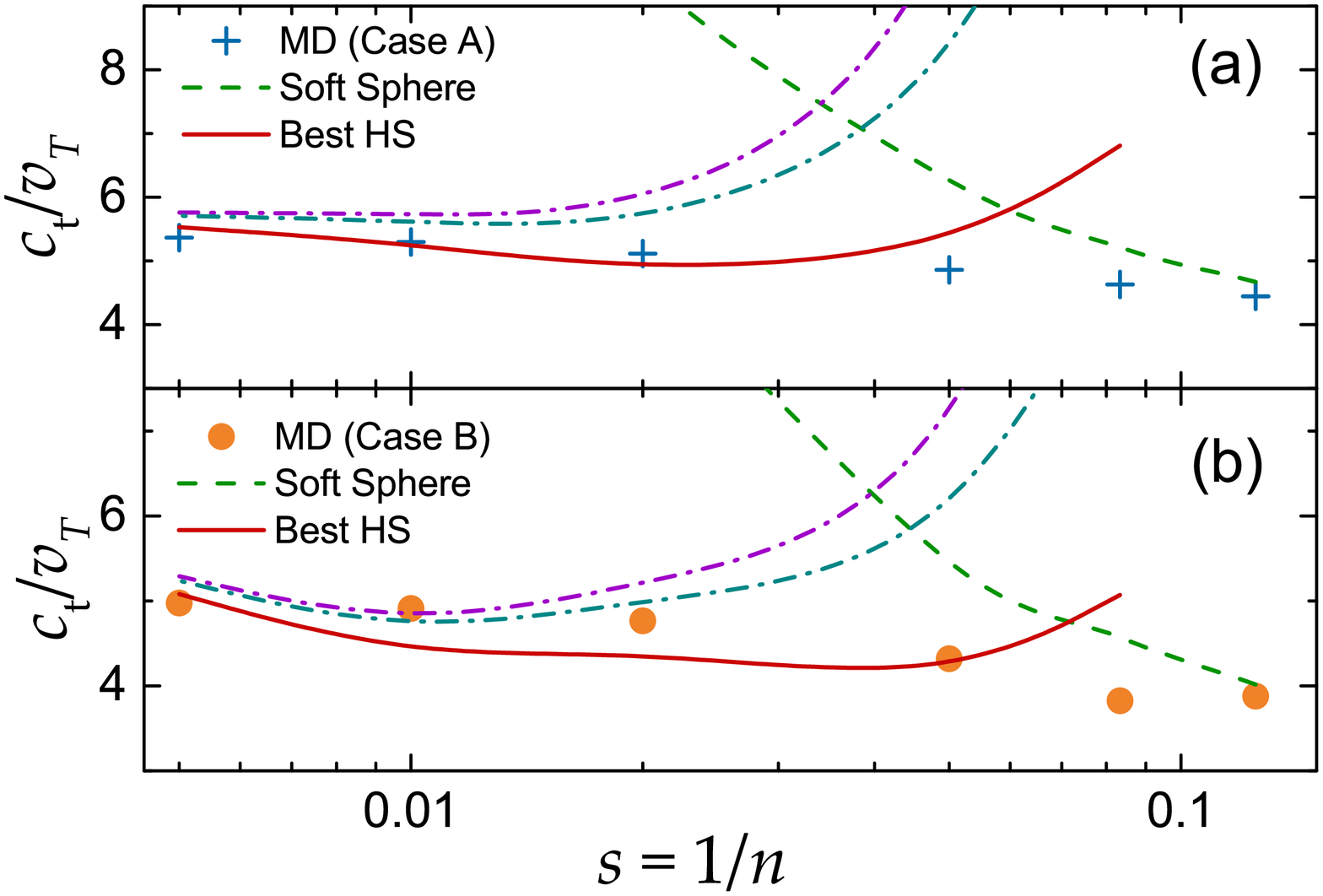}
\caption{Reduced transverse sound velocity, $c_t/v_{\mathrm{T}}$ versus the softness parameter of the IPL fluid. The HS asymptotes are calculated using Eq.~(\ref{HSL-eq7}). Other notation is the same as in Fig.~\ref{HSL-Fig3}.}
\label{HSL-Fig4}
\end{figure}

The theoretical and MD results are compared in Figs.~\ref{HSL-Fig3} and \ref{HSL-Fig4}, showing our main result. Here, the sound velocities expressed in units of the thermal velocity $v_{\rm T}=\sqrt{T/m}$, are plotted.
The dashed curves correspond to SSp (Zwanzig-Mountain) expressions, illustrating their range of suitability. The
SSp description fails at $n\simeq 20$, if judged from $c_s$ in Fig.~\ref{HSL-Fig3} and at $n\simeq 10$ if considering $c_t$ in Fig.~\ref{HSL-Fig4}. The red solid curves correspond to the HS asymptote with the effective HS diameter evaluated from Eq.~(\ref{HSL-eq8}) with ${\mathcal A}=(2/3)^{1/n}$ (other approaches converge to the same result at $n\rightarrow \infty$, but somewhat overestimate elastic moduli at finite $n$).  

%

\section{Discussion}

Our results can be useful to better understand a broad range of previous studies, suggest future research directions, and are not limited to model IPL fluids. In dense fluids, the motion of an atom  is dominated by repulsion from its nearest neighbors. Hence, the system properties are mainly governed by the shape of the interaction potential in a relatively narrow range of distances, near the average interparticle separation~\cite{KhrapakJCP2011,BohlingJPCM2013}.
In this region, the (extended) IPL potential can accurately fit the actual potential~\cite{BohlingJCP2014}.
The fit defines an effective IPL exponent $n_{\mathrm{eff}}$, regulating the effective softness of the actual interaction.



For example, the typical effective IPL exponent for the Lennard-Jones (LJ) potential, describing liquified noble gasses, is $n_{\mathrm{eff}}\simeq 18$ (at moderate densities)~\cite{PedersenPRL2008}:
This might appear to be surprising, but the reason is that the attractive ($\propto r^{-6}$) term of the LJ potential makes its repulsive short-range branch considerably steeper than just the $\propto r^{-12}$ repulsive term.
Only at high densities $n_{\mathrm{eff}}$ approaches 12~\cite{BohlingJCP2014}.
This should be kept in mind when analysing instantaneous elastic moduli (in particular, the shear modulus) in liquified noble gasses~\cite{KhrapakMolecules2020}.
The same obviously applies to generalized LJ $m$-$n$ potentials, in particular at $m>12$.

In many liquid metals, the interactions can be approximated by the IPL potential.
Extensive density functional theory (DFT) calculations of 58 liquid elements at their triple points demonstrate that most metallic elements exhibit strong correlations between virial and potential energy and thus obey ''hidden scale invariance'' even at these relatively low densities~\cite{HummelPRB2015}. The structure and phase diagrams of many such elements are consistent with the IPL model.
Typical DFT computed density scaling exponents yield $n_{\rm eff}\lesssim 10$. However, in some cases $n_{\rm eff}$ are quite large. For instance, for Rh, Cd, Os, Ir, and Pt we observe $n_{\rm eff}\gtrsim 15$ with an extreme value $n_{\rm eff}\simeq 23.7$ for Au~\cite{HummelPRB2015}. A careful account of elastic properties of these systems is warranted.

Recently, a microscopic model for the temperature dependence of the shear viscosity and fragile-strong behavior of liquid metals in the supercooled regime has been developed by combining the shoving model with the SSp expression for $G_{\infty}$~\cite{ChevallardPRR2020}. The repulsion steepness of the interaction potential has emerged as the crucial parameter governing the glass fragility, which has been shown to increase monotonously with the repulsion steepness. This result is heavily based on the SSp expression for $G_{\infty}$, and can be naturally tested with a more appropriate expression for $G_{\infty}$ in the regime of steep HS-like interactions.


\section{Conclusion}

To conclude, the comprehensive analysis of different approximations for classical liquids, considered from \emph{macroscopic} (elastic properties) and \emph{microscopic} (excitation spectra) points of view, clearly reveals when soft spheres become hard spheres.
The results have allowed us to unravel the paradoxical divergences of classical Zwanzig-Mountain formulas and to determine the range of their suitability, providing a useful input for future studies of fluids and glasses, from atomic to macromolecular systems.


\section*{Acknowledgement}
We would like to thank V. Nosenko and M. Schwabe for careful reading of the manuscript. N.P.K. and S.O.Y. are grateful to BMSTU State Assignment for infrastructural support. Analysis of the effect of interaction steepness on collective modes in fluids was supported by the Russian Science Foundation, Grant No. 20-12-00356.

\bibliography{HSL-Ref}

\end{document}